\documentstyle[aps,eqsecnum,prb,floats]{revtex}

\begin{document}
\draft
\twocolumn
\input epsf

\title{Extending holographic LEED to ordered small unit cell superstructures}
\author{K. Reuter \cite{Erlangen}, J.A. Vamvakas and D.K. Saldin}
\address{Department of Physics and Laboratory for Surface Studies,\\
University of Wisconsin-Milwaukee,\\
P.O. Box 413, Milwaukee, Wisconsin 53201,\\
U.S.A.}
\author{V. Blum, M. Ott, H. Wedler, R. D\"oll and K. Heinz}
\address{Lehrstuhl f\"ur Festk\"orperphysik, Universit\"at\\
Erlangen-N\"urnberg,\\
Staudtstr. 7, D-91058 Erlangen, Germany
\smallskip
\small
\begin{quote}
Following on the success of the recent application of holographic LEED to
the determination of the 3D atomic geometry of Si adatoms on a SiC(111) p(3$%
\times$3) surface, which enabled that structure to be solved, we show in
this paper that a similar technique allows the direct recovery of the local
geometry of adsorbates forming superstructures as small as p(2$\times$2),
even in the presence of a local substrate reconstruction.
\flushleft{PACS-numbers: 61.14.Nm, 61.10.Dp, 61.14.Hg, 68.35.Bs \hfill
{\bf Phys.\,Rev.\,B 58(5), (1998) {\em in press}}
}
\end{quote}
}
\maketitle
\narrowtext

\section{Introduction}

The short inelastic scattering length of low energy electrons makes them
ideal tools for surface crystallography. The energy variations of the
intensities (I(V) curves) of the reflected Bragg beams depend sensitively on
the structure of the few outermost atomic layers, which is usually deduced
by a trial-and-error fitting to simulated I(V) curves from model structures 
\cite{Pendry74}.

Even for a single model structure the calculation of a set of I(V) curves is
no trivial task. The calculation requires two main ingredients: first, a
representation of the scattering properties of each of the atoms of the
presumed structure - this is parametrized in a set of energy-dependent {\it %
phase shifts} for each angular momentum quantum number. Second, it is
necessary to evaluate the scattering of electrons amongst all atoms of the
sample. During the 1960's and 1970's many ingenious methods were developed
to efficiently sum the multiple-scattering paths followed by the electrons
from source to detector \cite{Pendry74,vanHove79}.

Although many of the simpler surface structures were solved during this
period, it was also becoming clear that more complicated structures would
probably forever lie beyond the reach of standard multiple-scattering
trial-and-error fitting methods due to the exponential scaling of computer
time with the number of parameters to be fitted. A simple example makes this
abundantly clear: suppose we need to determine the 3 Cartesian coordinates
of N symmetry inequivalent atoms, and for this purpose we consider models
with 10 values of each coordinate. The time required for an exhaustive
search amongst these parameters is $T=S\times 10^{3N}$, where $S$ is the
time required to calculate the LEED spectra from each model. Taking $S=1$
second and $N=3$ we find $T=30$ years. Just doubling the number of atoms to $%
N=6$ results in a time $T$ of the order of the age of the universe. In the
latter case, increases in computer speeds by even a million-fold would only
bring $T$ down to a few millenia! (Of course, the above time estimates
reduce considerably if it were judged that not all three Cartesian
coordinates need to be determined independently).

Obviously there is a need to examine some alternatives to exhaustive
trial-and-error searches. Consequently, procedures applying a directed
search in the multidimensional parameter space have been developed (for
reviews see e.g refs. \onlinecite{Vanhove93,Heinz95}). In this context the
testing of simulated annealing-like and genetic algorithms \cite{Rous93} are
certainly steps in a most promising direction. Also, quasidirect methods
were developed allowing a search-free optimization of structural parameters
once a model near the correct structure had been found or guessed \cite
{Pendry88}. The other approach that has shown considerable promise for LEED
in recent years has been the holographic method \cite{Szoke86}.

This has been applied primarily to surfaces containing adatoms of either the
same or different species than the surface atoms. Initially it was used to
reconstruct a 3D image of the local geometry of disordered atomic adsorbates
on a surface from the {\it diffuse} LEED (DLEED) patterns in which the
electron intensity between the Bragg spots is capable of a holographic
interpretation \cite{Saldin90}. Recent work has shown that essentially the
same numerical algorithm is able to form an image of the local geometry of
Si on a SiC(111)-p(3$\times$3) surface \cite{Reuter97_2}. This work was
important in that it showed that the holographic method can be applied to
the case of an ordered array of adatoms on a surface by operating on the
intensities of superstructure (or fractional-order) Bragg spots, and in that
it actually was the catalyst to the solution of this rather complex
structure that was most likely beyond the scope of the current techniques of
conventional LEED analysis.

As has been pointed out earlier \cite{Heinz91,Mendez92,Hu92,Tong92}, a set
of superstructure Bragg spots from an ordered array of adsorbates may be
thought of as sampling the corresponding diffuse LEED intensities at the
positions of these spots. In the case of a large adsorbate unit cell like
the (3$\times$3) one in the above cited case, the diffuse intensities may be
thought as being sampled over a rather dense grid in the reciprocal-space
planes parallel to the surface. However, within the class of ordered
adsorbate superstructures, those of smaller unit cells are much more
frequently observed. One such set of superstructures which still can present
difficulties for present state-of-the-art conventional LEED methods are the
p(2$\times$2) structures. It is the aim of the present paper to show that
even this class of adsorbate superstructures may also be brought within the
purview of holographic LEED methods.

\section{The 'CORRECT' algorithm for diffuse LEED}

Holographic methods have already been employed successfully to reconstruct
the local geometry of disordered atomic adsorbates on a crystal surface \cite
{Wei94,Heinz94,Saldin96,Saldin97}. In the corresponding diffuse LEED
patterns, diffraction intensity can be measured in the areas between the
Bragg spots. This intensity arises from elastic scattering from a surface
exhibiting structural elements lacking lateral translational symmetry. If
one considers elastic scattering from a surface containing a single
adsorbate atom, it is obvious that the DLEED intensity $H_{a}$ it gives rise
to can arise from just those scattering paths that include a scattering from
the adsorbate atom since this is the only one that breaks the lateral
periodicity of the surface. The diffuse intensity from a disordered layer of
such adsorbates in the same local sites of a flat substrate may be written 
\cite{Pendry84}:

\begin{equation}
H({\bf k}) = H_{a}({\bf k}) S({\bf k}_{\parallel})  \label{dleed}
\end{equation}

\noindent where ${\bf k}\equiv ({\bf k}_{\parallel },k_{\perp })$ is the
wavevector of the detected electron, ${\bf k}_{\parallel }$ its component
parallel to, $k_{\perp }$ that perpendicular to the surface, and $S$ is a
lattice factor that quantifies the degree of long-range order amongst the
adsorbates. In the case of perfect lattice-gas disorder, which might be
expected at very low adsorbate coverages, $S$ becomes constant and equal to $%
N$, the total number of adsorbates illuminated by the electron beam, except
in the parts of the diffraction pattern occupied by the substrate Bragg
spots (which are excluded in practice), where it is equal to $N^{2}$. Thus
the accessible part of the DLEED pattern is just a more intense version of $%
H_{a}$.

The holographic view of DLEED \cite{Saldin90} focuses on the fact that the
obligatory scattering at an adsorbate leads to a natural separation of all
scattering paths: electrons whose final scattering is by an adsorbate form
the {\em reference} wave $R({\bf k})$, while those scattered subsequently by
substrate atoms before reaching the detector provide the {\em object} wave $%
O({\bf k})$. On this picture of the adsorbate as a microscopic
beam-splitter, $H_{a}$ can be written:

\begin{equation}
H_{a}({\bf k}) = \left| R({\bf k}) + O({\bf k}) \right|^2  \label{holoint}
\end{equation}

For a single DLEED pattern, it was suggested that the holographic
reconstruction algorithm derived by Barton \cite{Barton88} for photoelectron
holography could be used to produce a 3D image of the local atomic
environment of the adsorbate. Consequently, the numerical inversion
algorithm can be formulated as a phased 2D Fourier transform of the DLEED
data over ${\bf k}_{\parallel}$.

Due to the fact that DLEED patterns of different electron energies can be
measured easily, the idea emerged that information from several such
patterns could profitably be combined in reconstructing the image \cite
{Saldin90}. Indeed, multi-energy reconstruction algorithms \cite
{Barton91,Tong91} led to a considerable improvement in the images that could
at that time be called reliable for the first time. Apart from suppressing
the unwanted holographic twin image, the additional stationary-phase
condition arising from the integral over energies was designed to single out
the contributions due to the kinematic object wave. A benefit from this is
that corrections for the remaining anisotropies of the reference and object
waves may then be performed by assuming rather simple forms for these
quantities \cite{Heinz93}. Subsequently, many variants of multi-energy
reconstruction algorithms have been suggested \cite{SaldinRev}.

The most successful of such holographic DLEED inversion algorithms to date
has been the {\em Compensated Object- and Referencewave Reconstruction by an
Energy-dependent Cartesian Transform }(CORRECT)\cite{Saldin95}. Unlike
previous multi-energy algorithms that perform the 3D holographic
reconstruction integral in a polar coordinate system (angle and energy),
this transform operates on a Cartesian data input (${\bf k_{\parallel }}%
,k_{\perp }$) \cite{Saldin97}. Considering that a superstructure spot is
characterized by a constant ${\bf k_{\parallel }}$ under changes of electron
energy, this algorithm has the advantage of being most suitable for an
extension to diffraction data from ordered superstructures. With this
algorithm, the reconstructed real space distribution around the adsorbate $%
\left| B({\bf r})\right| ^{2}$ (where ${\bf r}\equiv ({\bf r_{\parallel }}%
,z) $ is a position vector relative to an origin at an adsorbate with
components ${\bf r_{\parallel }}$ parallel and z perpendicular to the
surface) can be calculated {\em via}:

\begin{eqnarray}
B({\bf r}) &=& \int\!\!\!\int_{{\bf k}_{\parallel}} \left[ \int_{k_{\perp}}
K({\bf k}_{\parallel}, k_{\perp};{\bf r}) \chi({\bf k}_{\parallel},
k_{\perp}) e^{-(ikr-k_{\perp}z)} dk_{\perp} \right]  \nonumber \\
& & e^{i{\bf k}_{\parallel}\cdot{\bf r}_{\parallel}} d^2{\bf k}_{\parallel}.
\label{correct}
\end{eqnarray}

Apart from the obvious fact that this involves a 3D integral over reciprocal
space, two noteworthy features are: first, it operates not directly on the
measured intensities $H$, but rather on a contrast-enhancing and normalizing
function

\begin{equation}
\chi({\bf k}_{\parallel}, k_{\perp}) = \frac{H({\bf k}_{\parallel},
k_{\perp}) - H_{av}({\bf k}_{\parallel})}{H_{av}({\bf k}_{\parallel})}
\label{chi}
\end{equation}
with 
\begin{equation}
H_{av}({\bf k}_{\parallel}) = \frac{\int H({\bf k}_{\parallel}, k_{\perp})
dk_{\perp}}{\int dk_{\perp}}.  \label{hav}
\end{equation}

It has been shown theoretically \cite{Saldin95} that the use of such a $\chi$%
-function helps to partially remove the self-interference terms $\left| R(%
{\bf k}) \right|^2$ and $\left| O({\bf k}) \right|^2$ in the DLEED intensity
in Eq. (\ref{holoint}), which give rise to high intensities around the
origin of the reconstructed real space distribution. More importantly, $\chi$
has been designed to remove effects due to partial ordering amongst the
adsorbates. Since for flat substrates (i.e. not those exhibiting steps) the
lattice factor $S$ in Eq. (\ref{dleed}) is an exclusive function of ${\bf k}%
_{\parallel}$, it is unaffected by the integrals over $k_{\perp}$ in the
definition of $\chi$ and hence cancels out. Deviations from perfect
lattice-gas disorder at higher coverages, which give rise to modulations of $%
S$, can thus be efficiently suppressed \cite{Saldin97}.

The anisotropic scattering of low energy electrons by the adsorbate beam
splitter makes the local environment of the adsorbate show up only within
the forward-scattering cone of the incident wave when images are
reconstructed by previous inversion algorithms \cite{Wei92}. Full 3D images
are only formed with such algorithms by taking sets of DLEED data of
different electron energies for each of a number (at least two) of different
directions of incident electrons \cite{Wei94,Heinz94}. Of course, this
causes considerable experimental effort and so it would be much preferable
to use a set of DLEED patterns from just the most reliably-measured normal
incidence data which allow to improve the data quality by off-line averaging
according to the underlying surface symmetry. This becomes possible with the
use of the CORRECT algorithm, due to its compensation for the anisotropy of
the reference wave by estimating its value at the position of a scatterer,
by inclusion of the kernel

\begin{equation}
K({\bf k}_{\parallel}, k_{\perp}; {\bf r}) = \left[ \frac{f_a({\bf k}_i\cdot%
{\bf \hat{r}}) + C}{r} \right]^{-1}.  \label{kernel}
\end{equation}

\noindent under the integral in (\ref{correct}). Here $f_a({\bf k}_i\cdot%
{\bf \hat{r}})$ is the atomic scattering factor of the adsorbate, ${\bf \hat{%
k}}_i$ the direction of electron incidence, and $C$ the so called kernel
constant (which we take to be real), and which represents a zeroth-order
approximation to the backscattering by the substrate prior to scattering by
the adsorbate. An order of magnitude estimate can be made for $C$ from its
theoretical derivation (leading to $C_{ONi} \simeq 0.75${\AA} and $C_{KNi}
\simeq 2.20${\AA} for both the systems O/Ni and K/Ni treated in this paper) 
\cite{Saldin95}. In general, the effect of $C$ is well defined and hence its
value can also simply be optimized, such that all atoms in the local
adsorption geometry show up with similar brightness \cite{Reuter97_1}.

This algorithm has been shown to give reliable images using theoretical \cite
{Saldin95,Reuter97_1}, as well as experimental DLEED data \cite
{Saldin96,Saldin97}.

\section{DLEED and LEED intensities}

Even though research on surfaces giving rise to diffuse LEED intensities can
provide important information on initial (disordered) stages of adsorption,
the great majority of structures investigated manifest long-range order in
the plane parallel to the surface (henceforth referred to as the {\it lateral%
} plane). Particularly, many systems, where this order results in the
surface having larger lateral unit cells than that of the bulk, put rather
heavy demands on the standard quantitative LEED analysis. The large number
of structural parameters to be determined increases exponentially the number
of trial structures to be considered. For such structures, the lattice
factor $S$ becomes a sum of $\delta $-functions with peaks at the reciprocal
lattice rods due to the period of the new superstructure. In other words,
destructive interference between the waves originating from different
adsorbate-substrate clusters extinguishes all diffraction intensity but that
concentrating in the newly formed, sharp superstructure spots.

Nevertheless, these superstructure spots have been proven to contain the
same crystallographic information on the local environment of the adsorbate:
as long as scattering between different adsorbates can be neglected, the
diffuse intensity from an adsorbate and its local surroundings corresponding
to a particular value of ${\bf k}_{\parallel}$ has the same energy
dependence as the superstructure spot intensity from an ordered array of
such adsorbates in an equivalent local adsorption geometry \cite{Heinz91}
and of a surface reciprocal lattice vector equal to the same value of ${\bf k%
}_{\parallel}$. As pointed out earlier \cite{Mendez92}, the superstructure
spots may be thought of as sampling the DLEED intensity distribution of the
corresponding lattice gas on a finite grid. It was also shown that the
neglect of intra-adsorbate scattering holds even for relatively dense
superstructures under normal electron incidence \cite{Mendez92}.

With this understanding, a DLEED holographic inversion algorithm like
CORRECT may in principle be applied to superstructure spot intensities -- in
particular, since the latter takes its input diffraction intensities on a
Cartesian grid in reciprocal space. The only apparent difference is that the
finite sampling breaks down the integral over ${\bf k}_{\parallel}$ to a
discrete sum. However, a numerical implementation of the reconstruction
algorithm does this even in DLEED, so that the real difference is the
strongly {\em reduced} data resolution in ${\bf k}_{\parallel}$ \cite
{Reuter97_1}. In this respect, the larger the new surface period, the more
fractional order spots and hence the more data left for the algorithm and
the more similar the situation becomes to the former DLEED case. Recently,
the CORRECT algorithm was thus successfully applied to the p(3$\times$3)
reconstruction of SiC(111) \cite{Reuter97_2}. The high density of
superstructure spots corresponding to such a large reconstructed surface
unit cell, allowed the application of the DLEED algorithm without any
modifications caused by the finite sampling. Differences that appear for
small ordered superstructures with their even further restricted data base
will be addressed in the next section.

Note that the application to SiC(111) highlights the fact that for ordered
superstructures, it becomes irrelevant for holography, whether the
microscopic beam-splitter is an adsorbate on top of a substrate or belongs
intrinsically to the substrate itself as in the case of adatoms in surface
reconstructions. This increases considerably the number of systems to which
holographic LEED may be applied. The main requirements are that the
beam-splitter atom is the principal cause of the break of the bulk lateral
periodicity and that there be only one such atom per superstructure unit
cell (in order to prevent intermixing of images centered on several such
holographic reference-wave sources). If large, inequivalent adsorbate
overlayer domains exist on the surface the resulting image would be a simple
superposition of the local atomic structure around each adsorbate.
Since the reconstructed image reveals essentially the local bonding geometry
of the adsorbates, even the existence of more than one inequivalent
superstructure domain is immaterial, provided the local geometries of the
adsorption sites are identical.

As a last point, it is important to mention that for the application to
conventional LEED data, the strong reduction in ${\bf k}_{\parallel }$ data
resolution goes hand in hand with a strong increase in the available energy
range and data quality. Conventional LEED spot intensities can be measured
up to much higher energies ($\simeq $ 400-500eV and even higher at low
temperatures) than DLEED distributions, which suffer from disturbing effects
from the much brighter substrate Bragg spots \cite{Mendez92} and greater
contributions from thermal diffuse scattering at higher energies. Since
atomic scattering factors show a much weaker energy dependence at elevated
energies, this additionally accesible data range is also particularly
favourable for any holographic inversion algorithm, which tend to rely on
rather smoothly varying electron scattering properties. Further, the
signal-to-noise ratio of the bright discrete spots is much higher than that
of diffuse intensities, and contributions from thermal diffuse scattering
are less important and easier to subtract. In conclusion, the data
acquisition of discrete LEED diffraction spots is thus much more standard
and reliable than that of diffuse LEED distributions. Consequently, the
extension of holographic inversion algorithms to ordered systems will make
this technique much more applicable to a wider range of systems.

\section{Limits of validity for small superstructures}

Even though the above-mentioned application to SiC(111)-p(3$\times$3)
employed the unmodified CORRECT algorithm, it can only be seen as an
intermediary step towards the real extension of DLEED holography to ordered
superstructures. There are very few such large unit cell reconstructions
with the additional requirement of having only one prominent atom to serve
as a holographic beam-splitter. In general, surface structures will show
smaller periodicities with even less fractional-order spots remaining for
the inversion algorithm. In this case, the coarse sampling must evidently
result in perturbing effects on the integral transform designed for
continuous data distributions. Hence, there are two questions to be
addressed: (i) Will there be restrictions on the general validity of the
method? (ii) Is there a way to make more efficient use of the smaller
quantity of remaining data? We will treat question (i) first and defer
question (ii) to the next section.

When considering restrictions on the general validity, the key point to
notice is that the primary consequence of the prevailing surface periodicity
is to reduce the continuous ${\bf k}_{\parallel}$ integral in Eq. (\ref
{correct}) to a discrete sum. Since the term inside the square brackets in
Eq. (\ref{correct}) is a slowly varying function of ${\bf r}$, this
two-dimensional integral is essentially a Fourier transform in ${\bf k}%
_{\parallel}$. The problem encountered here is hence similar to the well
known numerical problem of discrete Fourier transforms. The sampling theorem 
\cite{recipes} states in this case, that when a function that is
not-bandwidth limited is sampled on discrete intervals of magnitude $\Delta$%
, there will be a frequency range in the dual space limited by the {\em %
Nyquist} frequency $f_c$ \cite{warning}:

\begin{equation}
\left[ -f_c ; f_c \right] \mbox{\quad with \quad} f_c = \frac{2\pi}{2\Delta}.
\label{critical}
\end{equation}

All power spectral density of the original function that would lie outside
of this range, will be spuriously moved inside by {\em aliasing}, and the
Fourier transform will be periodic due to aliasing of spectral density
outside of $f_c$.

For the application of this theorem to our present problem, consider for
simplicity a laterally one-dimensional example with a surface reconstruction
with a period of $n$ times the bulk one. This gives rise to $(n-1)$
fractional order diffraction spots per Brillouin zone in the LEED pattern.
The sampling interval $\Delta$ will thus be $g/n$, where $g$ is magnitude of
the reciprocal lattice vector corresponding to the lateral bulk periodicity
given by

\begin{equation}
g = \frac{2\pi}{a_p}
\end{equation}

\noindent and $a_p$ the lateral lattice constant. Substitution in (\ref
{critical}) determines the {\em Nyquist} critical frequency to be

\begin{equation}
f_c = \frac{n a_p}{2}.  \label{nyquist}
\end{equation}

Let us think of some hypothetical superstructure periodicities: $n=1$
represents same periodicity of bulk and surface, and hence results in no
fractional-order spots. Our holographic method is not applicable here. On
the other hand, $n>2$ leads to a larger number of fractional-order spots and
a larger usable data density. Any restrictions on the algorithm's validity
will be expected to affect such a case little, and one can argue that it
suffices to derive such limits for the worst case, namely $n=2$.

From (\ref{nyquist}) above, $f_c = a_p$ for this case. However, so far we
have neglected the fact that the integer order spots have to be excluded
from the inversion algorithm since they are dominated by pure substrate
scattering paths. This will double the effective sampling interval $\Delta$
to $g$ and consequently halve the critical frequency to $a_p/2$. On the
other hand, for real laterally two-dimensional cases, the effective sampling
interval may also lie between these two estimates, depending on the
periodicity in the other lateral direction. For example, consider the LEED
pattern from a p(2$\times$2) superstructure: there will be rows of
superstructure Bragg spots in the direction of each reciprocal lattice
vector at intervals of $g/2$ and alternate rows that include integer-order
spots where the superstructure spots are found at intervals of $g$. {\em Cum
grano salis} we may therefore argue, that the critical frequency for this
case must lie somewhere in the range

\begin{equation}
\frac{a_p}{2} < f_c < a_p.
\end{equation}

The sampling theorem now implies that any real space intensity represented
by the function $\left| B({\bf r})\right| ^{2}$ in Eq. (\ref{correct}) that
would normally appear outside a lateral range of $\pm f_{c}$, is moved
inside by aliasing. A natural consequence would be severe image distortions
due to these contributions. Fortunately, the algorithm is saved by the heavy
damping of the low energy electrons inside the crystal, which reduces
strongly any contribution to the diffraction intensity from atoms farther
from the adsorbate. Also, holographic fringes due to more distant atoms will
be of higher frequency in reciprocal space that may be beyond the available
data resolution. Thus we may deduce that any such aliased contributions
inside the critical frequency range $\pm f_{c}$ be small compared to the
real intensities corresponding to true atomic positions (in Fourier
transform terms this can be rephrased that $H({\bf k})$ is bandwidth
limited). This reasoning is corroborated by the fact, that in all previous
investigations using DLEED data, where no such aliasing effects are
expected, the only atomic positions recovered belonged to the direct local
adsorption geometry and no intensity due to further outlying atoms appeared
in the reconstructed images \cite{Saldin96,Saldin97,Saldin95,Reuter97_1}.

The more important consequence of the sampling theorem in the present case
is the folding out of intensity outside of the critical range $f_c$. In
other words, aliasing restricts the range of lateral validity of the
algorithm to

\begin{equation}
\frac{a_p}{2} < f_c < a_p
\end{equation}

\noindent for p(2$\times$2) structures. Outside of this range one cannot be
sure whether reconstructed intensity is due to real atoms or solely due to
aliased contributions. Note, that even in this worst case, the aliasing does 
{\em not} destroy the algorithm's applicability: the latters' main task is
to recover atomic positions of the local adsorption geometry, or the local
adatom environment. In almost any imaginable case the lateral distance of
such neighbouring atoms is less than $a_p$, and hence still accesible to the
holographic inversion algorithm. In particular, for the case of a
hollow-site adsorption on a fcc(001) surface, which we discuss later, the
lateral distance of the 4 nearest first layer substrate atoms is $a_p/2$ and
thus well within the critical range. Even though the very restricted data
resolution will thus not lead to a breakdown of our holographic scheme, we
must keep clearly in mind its limited range in the direction parallel to the
surface.

\section{Dropping the $\chi$-function}

We now consider the question whether the limited quantity of remaining data
in conventional LEED should be treated differently by a holographic
algorithm. After establishing the equivalence, for our purposes, of DLEED
and LEED intensities, it is obvious that the general form of the 3D integral
transform does not need to be changed, since the algorithm still targets the
same scattering-path phases. On the other hand, as mentioned earlier, the
primary purpose for the design of the contrast-enhancing $\chi$-function in
Eq. (\ref{chi}) has been to suppress effects of partial ordering amongst the
adsorbates, which leads to modulations in the lattice factor $S$. However,
for the case of an ideally ordered superstructure on a flat surface, $S$
becomes a sum of $\delta$-functions with equal value at each of the
remaining reciprocal lattice rods. Any filtering effect of $\chi$ is thus no
longer needed.

The second advantage of presenting data to the inversion algorithm in the
form of the $\chi$-function has been the partial removal of the
self-interference terms in Eq. (\ref{holoint}) which would be expected to
give rise to just low-frequency oscillations in the diffraction pattern. The
principal argument for this is that oscillations in $H({\bf k})$ are due
mainly to the reference-object interference terms $R^{\ast}({\bf k})O({\bf k}%
)$ and $R({\bf k})O^{\ast}({\bf k})$ and that hence $H_{av}({\bf k}%
_{\parallel})$ as defined in Eq. (\ref{hav}) represents a rough
approximation to the remaining self-interference terms in $H({\bf k})$ \cite
{Saldin95}. It is clear, that this crude estimate causes a degradation of
the input diffraction data in the construction of the $\chi$-function. This
is of no importance in the DLEED case, where an abundance of data tends to
average out such deleterious effects, and allows predominantly the positive
consequences of the $\chi$-function to be noticed \cite{Saldin97}.

With the filtering effect not required for superstructures, and knowing
that, due to the limited quantities of available conventional-LEED data, it
will be advantageous to use rather the un-degraded measured intensities $H(%
{\bf k})$ as input to the reconstruction algorithm, we seek an alternative
way to suppress the self-interference contributions to the images. This can
actually be achieved with a much simpler procedure: since these terms in the
diffraction intensity are primarily of low-frequency in the diffraction
data, they would be expected to cause distortions mainly in the part of the
reconstructed image near its origin \cite{Saldin90,Hu92,Barton88,Saldin95}.
Given that the origin is defined by the position of the atomic
beam-splitter, no other atoms are expected within a hard-sphere radius of
that adatom or adsorbate. We therefore suggest, that ignoring any image
features within such a radius (i.e. setting $B({\bf r})$ to zero in that
region), is as effective a method of suppressing the self-interference terms
as by the use of a $\chi$-function, but without the danger of degradation of
the useful holographic fringes in the diffraction data.

\section{Test case of O/N\lowercase{i(001)} \lowercase{p}(2\lowercase{x}2)}

With these modifications, our algorithm should be applicable to LEED data
from smaller ordered superstructures. Such data are characterized by a lower
reciprocal-space density of fractional-order spots, and there may indeed be
a lower limit to this density for effective holographic reconstruction. Thus
in the order of surface structures, as characterized by overlayer sizes,
holographic LEED promises to occupy a niche essentially complementary to
that of conventional quantitative LEED methods: the smaller surface unit
cells possess a limited number of free structural parameters, and can be
almost routinely solved by standard trial-and-error procedures \cite{Heinz95}%
, while the larger superstructures are more problematical due to the
exponential scaling with the number of parameters, as discussed earlier. In
contrast, the larger density of fractional-order Bragg spots from the larger
overlayer unit cells makes their LEED pattterns more favorable for an
holographic analysis, as our earlier work on the SiC(111) (3$\times $3) has
already demonstrated \cite{Reuter97_2}. In the present paper we show that
even superstructures as small as p(2$\times $2), which have in the past
posed considerable difficulties for conventional LEED analysis, may be
brought within the purview of the holographic method.

As an initial test system we chose the p(2$\times $2) phase of O/Ni(001).
Because this is an extensively-investigated and well-known system, it is
appropriate for testing methodologic advances: since Ni is a strong
scatterer and O a weak one adsorbed very close to the surface, it can be
argued that this structure represents a rather unfavourable scenario for
holographic techniques, which rely on dominant scattering by the O atom to
enhance the reference wave and a suppression of intra-Ni and O-Ni multiple
scattering. A successful recovery of the adsorption site in this case would
suggest a general applicability of our method, especially on e.g. lighter
scattering substrates like Si.

\section{Image display scheme}

The reconstructed images in this paper will be displayed in perspective to
reveal their full 3D real space geometry. The function $\left| B({\bf r})
\right|^2$ is calculated from Eq. (\ref{correct}) on a grid of $0.2${\AA}
resolution inside a cylinder of depth 3.5{\AA} and a lateral radius of 2.0{%
\AA}, consistent with our estimate of the limits of lateral validity for the
unit cell parameter $a_p=2.49${\AA} for Ni. All atoms that specify the local
adsorption geometry of O/Ni(001)-p(2x2) are located within this volume and
their positions should therefore be retrievable by our method. Small
spheres, with diameters proportional to the reconstructed intensity $\left|
B({\bf r}) \right|^2$ and whose gray shading varies from white for the lower
intensities to black for the higher ones, are drawn at the grid points. The
origin of the coordinate system is defined by the adsorbate (beam-splitter),
which is artificially added to facilitate the understanding. No intensity is
calculated inside a sphere of 1.0{\AA} around this origin to suppress the
self-interference terms as reasoned above.

Even though the commonly used 2D cut-planes through a reconstructed image
may be more suited for an exact analysis of the atomic positions \cite
{Reuter97_1}, we believe that the overall 3D presentation of the
reconstruction result not only is more honest (no choice of presentation of
``particular'' cutplanes), but also focuses better on the main task of the
holographic technique: a direct insight into the essential features of a
structure, i.e. in this case the adsorption site.

\section{Simulated data from unreconstructed substrates}

As a first step we simulated LEED I(V)-data for a simplified
O/Ni(001)-p(2x2) geometry: O residing in the four-fold symmetric hollow site
of a bulk-truncated Ni(001) surface at an adsorption height of 0.9{\AA} \cite
{Oed90}. The simulation was carried out with the standard van Hove-Tong
computer code \cite{vanHove79} using up to 11 phase shifts representing
atomic scattering properties (this was found to be appropriate for the
envisioned energy range). The data set generated comprises the I(V)-curves
of all fractional order spots for the energies 100-400 eV. This represents a
typical and experimentally feasible range for LEED investigations. To
closely simulate the conditions of a typical LEED experiment, the only data
used was that collectible from within a $50^{o}$ polar semi-angle, which
corresponds to the standard angular range of conventional electron detectors 
\cite{Heinz95}, and additionally had been shown to be the most appropriate
in earlier holographic DLEED investigations \cite{Reuter97_1}.

The reconstructed image shown in Fig. 1 clearly identifies the four-fold
adsorption site. An unambiguous distinction between the 5 atoms of the Ni
local adsorption geometry and remaining residual ``noise'' on the image is
easily possible. The highest such noise at non-atomic locations is at 44\%
of the intensity of the identified atoms. The constant $C$ in the integral
kernel $K$ of Eq. (\ref{kernel}) has been optimized to a value of 0.70{\AA}
to give approximately equal intensity for both 1st and 2nd layer atoms \cite
{Reuter97_1}. This value agrees very well with the order-of-magnitude
estimate derived from theory as described in section II.

The most important property of a direct inversion algorithm when applied to 
{\em a priori} unknown systems, is its stability. No fundamentally different
information, that could lead to wrong predictions of structural features,
should be reconstructed when changing the parameters of the algorithm. With
most of the data base parameters dictated by the experiment, the most
influential one left to be adjusted is the energy range of the data. Since
the effect of the finite energy integral in Eq. (\ref{correct}) is only to 
{\em suppress}, and not to {\em erase} multiple scattering contributions, it
is to be expected that the image quality can vary as a function of the
energy range used. Due to anisotropies in the atomic scattering factors
sometimes even smaller ranges may lead to improved images. However, changing
the upper and lower energy boundaries should not result in completely
different images, on the basis of which wrong adsorption sites would be
infered. We checked on this and found the image to be stable when increasing
the lower boundary or decreasing the upper one by approximately 60eV, i.e.
the maximum intensities were always found at the site of the 5 identified
atoms. The use of even smaller energy ranges caused the collapse of the
algorithm as expressed by very noisy images with many new intensity peaks.
So, no false adsorption site could have been obtained from the present data.

\begin{figure}[tbp]
\epsfxsize=0.5\textwidth \epsfbox{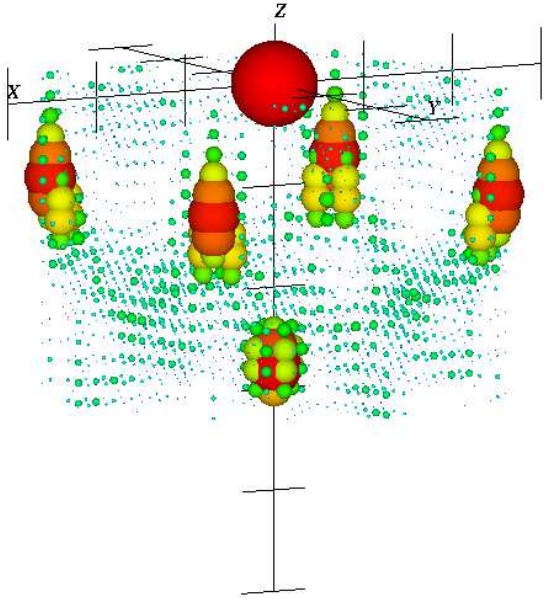}
\caption{Recovered local geometry of the O adsorption site from LEED data
calculated for a model Ni(001)-p(2$\times$2) surface with an unreconstructed
substrate. The electron energy range of the data used is 130-400 eV, the
kernel constant $C = 0.70${\AA} and the maximum noise level in the image was
found to be 44\% of the maxima denoting the atom positions. For details on
the display procedure, see Section VI.}
\end{figure}

The vertical position of all atoms is retrieved to within a 0.2{\AA}
accuracy. As already experienced with DLEED data \cite{Reuter97_1}, the
lateral position of the 1st layer Ni atoms, that appear only due to the
inclusion of the integral kernel $K$, has a much greater uncertainty of $%
\simeq 0.5${\AA}. Also, this lateral position can shift inwards or outwards
(see results in the next section) depending on the exact scattering
geometry. This effect has to be related to the simplicity of the
compensating kernel, which at its present stage includes only a zeroth order
approximation to the backscattering properties of the system \cite{Saldin95}
and is hence very sensitive to details of the atomic scattering factors.
However, the unambiguous recovery of the essential features of a structure
(like the adsorption site), is more than sufficient to allow a subsequent
refinement by the conventional quantitative LEED analysis.

\begin{figure}[tbp]
\epsfxsize=0.5\textwidth \epsfbox{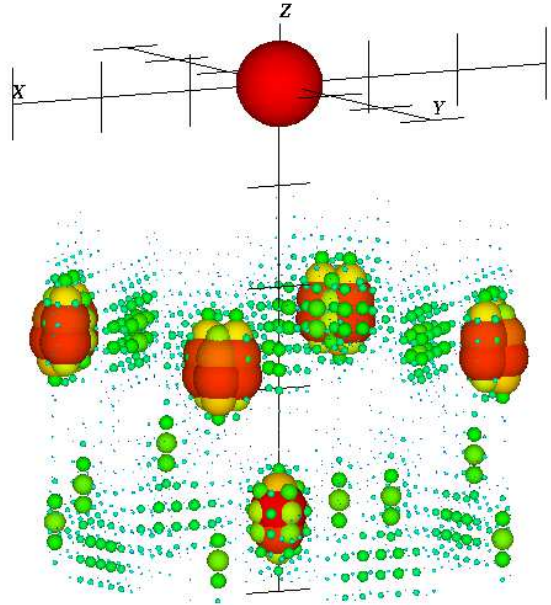}
\caption{Same as Fig. 1, except that the LEED data is calculated from a
model that had been fitted to experimental LEED data in a previous
quantitative LEED study. This model includes reconstruction of the substrate
and fitted thermal vibration parameters. The electron energy range used was
112.5-362.5eV, the kernel constant $C = 0.95${\AA} and maximum noise level
found to be 37\% of the maxima at the atoms.}
\end{figure}

To verify the generality of the result obtained, we simulated a second data
set, with a different adsorbed chemical species and adsorption height, but
with all other parameters unchanged. Consistent with the determination of
the related K/Ni(001)-c(4$\times$2) structure \cite{Wedler93}, we positioned
K in the four-fold hollow site with an adsorption height of 2.56{\AA} in a
hypothetical p(2x2) arrangement. Fig. 2 shows that this adsorption geometry
is reconstructed with comparable quality and clarity as in the case of O/Ni.
The optimized kernel constant $C = 2.35${\AA} again corresponds very well to
its theoretical estimate. The slightly higher noise level of 63\% is
probably related to the K scattering factor, but nevertheless allows an
unambiguous identification of all 5 atoms \cite{size}. The new vertical
positions due to the enlarged adsorption height are again correct within 0.2{%
\AA} and the lateral shift of the 1st layer atoms is comparable to the
previous case.

The stability encountered with respect to changes of the energy range used
is even increased with respect to O/Ni (variations of up to 90eV are
possible), which corroborates our hypothesis of O/Ni as an unfavourable
system for holography: the smaller multiple K-Ni scattering due to the
increased adsorption height together with the stronger reference wave due to
the K, reduce the required minimum energy range for multiple scattering
suppression. Consequently, we are led to believe that the two results
presented here suggest the general applicability of the modified CORRECT
algorithm.

\begin{figure}[tbp]
\epsfxsize=0.5\textwidth \epsfbox{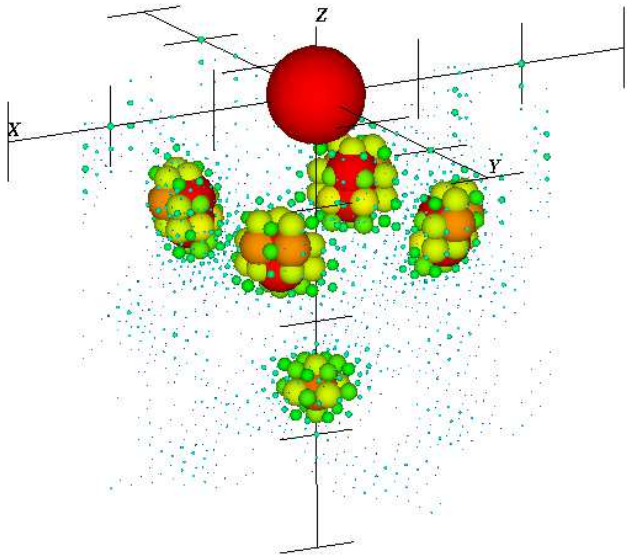}
\caption{Same as Fig. 1, except that the LEED data is calculated from a
model that had been fitted to experimental LEED data in a previous
quantitative LEED study. This model includes reconstruction of the substrate
and fitted thermal vibration parameters. The electron energy range used was
112.5-362.5eV, the kernel constant $C = 0.95${\AA} and maximum noise level
found to be 37\% of the maxima at the atoms.}
\end{figure}

\section{Realistic simulation from reconstructed substrates}

A typical concern often raised about results of the type presented in the
last section is that such simplified simulations cannot be compared to real
data as obtained from any realistic structure. Since the van Hove-Tong
computer code takes proper account of all multiple scattering, the main
difference between data as in the last section and theoretical data as used
for actual quantitative LEED analyses consists of two points: thermal
vibrations and structural deviations from bulk-like truncation like e.g.
layer relaxations or buckling. It is then argued that both effects might
possibly influence the holographic algorithm.

In order to address this question, we simulated a data set, where all
structural and non-structural parameters involved in the dynamical LEED
calculation were taken exactly as they had been obtained in the previous
quantitative analysis of the O/Ni(001)-p(2$\times$2) system \cite{Oed90}.
Coincidentally, our previous test system proved again to be a rather
unfavourable one with respect to both of the above-mentioned features: not
only is the determined vibrational amplitude of the O of 0.3{\AA} unusually
high, but also the Ni substrate experiences relaxations of the first two
layer spacings ($d_{12} = 1.80${\AA} and $d_{23} = 1.75${\AA}), and a
relatively strong buckling of 0.10{\AA} in the second layer. In common with
the previous simulations, the full data-set comprised again of all
fractional-order spots in the energy range 100-400 eV and the holographic
reconstruction was carried out by exactly the same procedure as described in
the last section.

Consequently, the result shown in Fig. 3 is particularly gratifying. The
local adsorption geometry is recovered just as clearly as from the
simplified data. Stability with respect to changes of the energy range used
and the accuracy of the recovered atomic positions are found to match very
well with the ones described in the preceding section. Note, however, that
the unstable lateral position of the 1st layer atoms now appears shifted
inward from the correct value as had already been pointed out in the last
section. Interestingly, the overall noise level in the image is with 37\%
even slightly lower for this ``tougher'' more realistic model. Knowing that
thermal vibrations lead to a reduction of the effective electron coherence
length, it might actually be argued that this effect helps reduce disturbing
multiple scattering contributions due to atoms further from the adsorbate
and hence benefits a holographic inversion.

However, the theoretical objection concerning structural deviations from
bulk-like positions has to be taken more seriously. Any substrate atom whose
position deviates from that expected from its ideal lattice position, like
the buckled 2nd layer Ni atom, could in principle act as another conduit for
electrons to the fractional-order Bragg spots. The standard argument for
neglecting these contributions in the holographic theory has been that atoms
that only differ slightly from their lattice positions contribute rather
weakly to the resulting diffraction intensity, in comparison to the major
rupture of the periodicity due to the introduction of a new atom, such as an
adsorbate or adatom. As a further theoretical test, we repeated the
simulation for the realistic O/Ni(001)-p(2x2) geometry, but this time with
the 2nd layer buckling artificially doubled to 0.2{\AA}. Even for this case,
the complete adsorption geometry was again retrievable under otherwise
identical parameters. The only apparent effect of the increased buckling
amplitude seemed reflected in a rise of the overall noise level to 62\%.
This can be seen as a corroboration of the argument that the breaking of the
bulk periodicity by a deviation of the position of a substrate atom leads to
contributions to the fractional-order intensities that may degrade the
simple picture of the adsorbate or adatom as the only path to those
fractional-order spots. However, the recovery of a good image even in this
case, clearly suggests that their contribution is small and only serves to
increase the background noise level on the image.

These results also help explain the success of our earlier work \cite
{Reuter97_2} for the reconstructed SiC(111)-p(3$\times$3) structure. In that
case the application of our algorithm to fractional-order Bragg spot data
was able to recover an accurate image of the local site of an adatom despite
the fact that other atoms in the surface unit cell were deviated from their
positions in an unreconstructed layer.

\begin{figure}[tbp]
\epsfxsize=0.5\textwidth \epsfbox{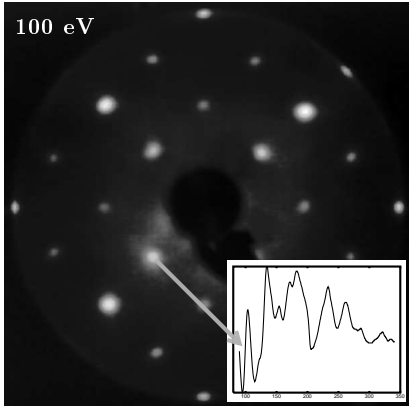}
\caption{Experimental LEED pattern from a O/Ni(001)-p(2$\times$2) surface
for an electron energy of 100 eV. The inset shows the I(V)-curve of the ($%
\frac{1}{2}$ 0) beam.}
\end{figure}

\section{Experimental data}

As a last step we try to invert experimental data of the O/Ni(001)-p(2$%
\times $2) system. After standard crystal preparation, the Ni substrate was
subjected to an exposure $2\cdot10^{-8}$ mbar of O for 60sec at 90 K.
Subsequent annealing at 500 K led to the formation of a sharp p(2$\times$2)
superstructure pattern as evidenced in Fig. 4. The measurement was performed
using the standard Video-LEED system developed in Erlangen \cite
{Heinz95,Mueller85} and included 4-fold symmetry averaging according to the
expected rotational symmetry of the diffraction pattern. The full data set
comprises the 8 symmetry inequivalent fractional-order beams closest to
specular reflection in the energy range 90-344 eV. Good agreement of this
enlarged data set was found with the three fractional-order spot I(V)-curves
used in the earlier quantitative LEED study \cite{Oed90}.

\begin{figure}[tbp]
\epsfxsize=0.5\textwidth \epsfbox{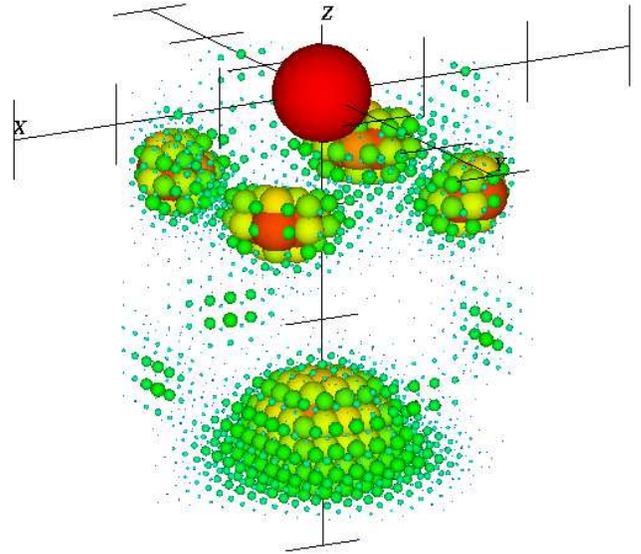}
\caption{Same as Fig. 3, except that the image is computed from experimental
data from a O/Ni(001)-p(2$\times$2) surface. The energy range used was
90-344 eV, the kernel constant $C = 0.33${\AA} and the maximum noise level
was 50\% of the maxima at the atoms.}
\end{figure}

Fig. 5 shows the reconstructed local adsorption geometry. Although the
experimental data set is smaller with respect to the total energy range and
the total number of fractional-order beams, we nevertheless find essentially
the same stability, accuracy and unambiguity of the image as described for
the other data sets. Even on reducing the number of beams used to the 5
nearest to the (00) beam in the diffraction patterns, the image continued to
reliably show the 4-fold hollow adsorption site with an only slightly
degraded overall image quality.

This final result shows, that no particularly large data set needs to be
measured to ensure a proper working of the reconstruction algorithm. Rather,
it is the same I(V)-curves that are measured for a quantitative LEED
analysis, that also provide the input to the holographic inversion.
Ultimately, it is only this similarity of required data that can make
holographic LEED a practically useful complement to the established
quantitative LEED analysis at the present time.

\section{Conclusions}

We have shown in this paper that a fully three-dimensional image of the
adsorption site of atoms forming a p(2$\times $2) overlayer on a metal
surface may be recovered from a data set that might typically be measured in
the usual practice of conventional LEED studies, i.e. one from just the
most-reliably measured normal incidence LEED data. We have demonstrated this
first for a simulated data set from models of O/Ni(001)-p(2$\times $2) and
K/Ni(001)-p(2$\times $2) surfaces with unreconstructed substrates. The usual
picture of holographic LEED from adsorbates on surfaces assumes that the
intensities of the fractional-order spots arise only from scattering paths
that include a scattering at the adsorbate. This is not strictly true in the
presence of substrate reconstructions which also cause deviations from the
periodicity of the bulk lattice. O/Ni(001)-p(2$\times $2) is a particularly
well-known system with an adsorbate-induced substrate reconstruction. We
show that realistic LEED I(V)-curves calculated from the accepted model of
this substrate reconstruction also give a clear and unambiguous holographic
image of the local adsorption site of the O atom. A similar image is also
reconstructed from experimental LEED data from this surface.

\section*{Acknowledgements}

K.R. is grateful for a DAAD travel grant enabling the stay at the University
of Wisconsin-Milwaukee during which this paper was written. The Milwaukee
group acknowledges support from the U.S. NSF (Grant No. DMR-9320275) and the
Erlangen authors support by the German DFG.


\begin{references}
\bibitem{warning}  Note, that Ref. \onlinecite{recipes} defines Fourier
transforms with the factor $2\pi $ separated from the frequency in the phase
factor. This factor does hence not appear in their formulation of the
sampling theorem.
\bibitem{size}  The depth of the reconstructed real-space volume was for
this calculation increased to $5.0${\AA } to include the 2nd layer atom.
\end{references}
\end{document}